\documentstyle[12pt]{article}

\def\bbar{\overline{\beta}}
\def\12{\frac{1}{2}}
\def\abar{\overline{\alpha}}
\def\tk{\tilde{k}}

\def\td{\tilde{\delta}}
\def\vbar{\overline{v}}
\def\tL{\tilde{\Lambda}}
\def\tb{\tilde{\beta}}

\def\tdelbar{\tilde{\partial}}
\def\kbar{\overline{k}}
\def\ttheta{\tilde{\theta}}
\def\tmu{\tilde{\mu}}
\def\tnu{\tilde{\nu}}

\def\diag{\mbox{diag}}
\def\tK{\tilde{K}}
\def \be {\begin{equation} }
\def \ee {\end{equation}  }
\def\thetbar{\overline{\theta}}
\def\Dbar{\overline{D}}
\def\Ubar{\overline{U}}
\def\ubar{\overline{u}}
\def\Vbar{\overline{V}}
\def\vbar{\overline{v}}

\def\tildar{\tilde{r}}

\def\jj{\overline{j}}
\def\tphi{\tilde{\varphi}}

\font\mybb=msbm10 at 12pt
\def\bbbb#1{\hbox{\mybb#1}}

\def\R {\bbbb{R}}

\begin{document}

\begin{titlepage}

\begin{flushright}
QMW-PH-99-18\\
AEI-1999-001\\
hep-th/9907046
\end{flushright}

\vspace{3cm}

\begin{center}

{\large \bf{The Geometry of Sigma-Models with Twisted Supersymmetry}}

\vspace{.7cm}

Mohab Abou Zeid\footnotemark[1] and Christopher M.\ Hull\footnotemark[2]

\vspace{.7cm}

\footnotemark[1] {\em Max-Planck-Institut f\"{u}r Gravitationsphysik,
Albert-Einstein-Institut,\\ Am M\"{u}hlenberg 5, D-14476 Golm, Germany}
\vspace{.7cm}

\footnotemark[2] {\em Physics Department, Queen Mary and Westfield College, \\
Mile End Road, London E1 4NS, U.\ K.\ }

\vspace{2.5cm}

June 1999


\vspace{1cm}

\begin{abstract}

We investigate  the relation  between
supersymmetry and  geometry for
two dimensional sigma models with target spaces of arbitrary signature, and
Lorentzian or Euclidean
world-sheets. In particular, we consider
  twisted forms of the two-dimensional  $(p,q)$ supersymmetry algebra.
Superspace formulations
of the $(p,q)$ heterotic sigma-models with twisted or untwisted supersymmetry
are given. For the
twisted (2,1)  and the pseudo-K\"{a}hler sigma models, we give extended
superspace formulations.

\end{abstract}

\end{center}

\end{titlepage}

\section{Introduction}

In~\cite{H1},  the analysis of~\cite{HW,H3,H2} on the geometry of  (1,0) and
(1,1) supersymmetric sigma-models was generalised to the case in which the
target space had arbitrary
signature, and
the conditions for   the theory to be invariant under extra
supersymmetries were
investigated.
Covariantly constant complex structures, i.e. (1,1) tensors $J$ satisfying
$J^2=-1$, led to extra
supersymmetries, each satisfying the usual superalgebra $Q^2 \sim P$, while
covariantly constant
real structures, i.e. (1,1) tensors $S$ satisfying $S^2= 1$, led to extra {\it
twisted} supersymmetries ~\cite{H1}, each satisfying the twisted superalgebra
$Q^2 \sim
-P$.
The number of structures of either type depended on the target space holonomy
of a certain connection
which had torsion if the sigma-model had a Wess-Zumino term.
For example, if the holonomy is contained in $USp(2m)$, there are three complex
structures $I,J,K$
satisfying the quaternion algebra with $SU(2)$
commutation
relations, while if
the holonomy is contained in $Sp(2m,\R)$, there is one complex structure  $J$
and two real structures $S,T$
satisfying the pseudo-quaternion algebra with $SU(1,1)$
commutation
relations.
The aim of this paper is to give the superspace formulation of these models
and to investigate their structure further.

Extended world-sheet supersymmetries have had two different uses in string
theory.
In the study of heterotic or type II strings, complex manifolds
such as Calabi-Yau spaces have played an important role. In these cases, the
string theory only
has gauged (1,1) or
(1,0) world-sheet supersymmetry, but the (1,1) or
(1,0) sigma-model on a suitable background can have extra rigid world-sheet
supersymmetries; on a
K\"{a}hler manifold, for example, $N=1$ world-sheet supersymmmetry is extended
to
$N=2$, and
$N=2$ superconformal field theory has played a central role in the study of
such
compactifications.
There are also string theories in which an extended world-sheet supersymmetry
is gauged, such as
those with $N=2$ local world-sheet supersymmetry, and in these a target space
with
either 4 Euclidean dimensions, or with 2 space and 2 time dimensions naturally
arises.
Our results on general signature have applications to both
 heterotic or type II strings in general signature~\cite{CMHnew,CMHnew2} and
to $(2,p)$ strings in
2+2 dimensions~\cite{forza,OV1,OV2}.

In~\cite{CMHnew,CMHnew2}, new  string theories were found in which the
10-dimensional
space-time had arbitrary
signature, and in some cases the world-sheet was Lorentzian (signature 1+1),
while in others it was
Euclidean  (signature 2+0). All of these were linked to the usual string
theories with target space
signature 9+1 and Lorentzian world-sheets by chains of
dualities~\cite{CMHnew,CMHnew2}. The
world-sheet formulation of
these string theories is  a sigma-model with target space of the appropriate
signature. Target
spaces that admit extra supersymmetries play an important role in the study of
solutions
of these theories, just as in the case of compactifying on Euclidean signature
internal spaces.

Another context in which non-Lorentzian signature target spaces have played a
role is in
$N=2$ strings, or more generally in strings with
$(2,0),(2,1) $ or $(2,2)$ world-sheet supersymmetry.
In these theories, the target spaces had signature 2+2 (or 4+0), and the
heterotic
theories were reduced (via a null reduction) to ones
with signature 1+1 or 2+1.
The (2,1) string is
of particular interest. It was shown by
 Kutasov and Martinec~\cite{KM1}, and by the same authors with
O'Loughlin~\cite{KM2}, that
  different vacua of
the    (2,1) superstring
  describe  the
$D1$-string  or the $D2$-brane, and
via dualities these are linked
 to all usual types of
ten-dimensional superstrings and to the eleven-dimensional
supermembrane~\cite{KM1,KM2}. This led to the suggestion  that the (2,1)
heterotic string
may provide many of the degrees of freedom of M theory, although this
approach has so far only yielded specially symmetric points in the moduli
space of vacua of M theory~\cite{M1,KM3}.
Martinec~\cite{M2,M3,M4} has proposed an interpretation of the (2,1) string
as describing the continuum limit of the matrix model of M theory~\cite{BFSS}
with all spatial dimensions compactified.

The (2,1) heterotic string~\cite{OV2} has a
four-dimensional target spacetime with signature (2,2)
that is required to have   an
isometry generated by a null Killing vector, which must be gauged.
In general there are obstructions to the gauging of a given
isometry~\cite{AH1,AH2}, and
the isometry is
required to be one for which these are absent. For
(2,2) signature, this null reduction yields either a space with signature
(2,1) (corresponding to
a membrane worldvolume~\cite{KM1,KM2}) or a space with signature (1,1)
(corresponding to a string worldsheet~\cite{KM1,KM2}). The theory defined
on a spacetime with signature (2,2) before null reduction is a theory of
self-dual
gravity with torsion coupled to self-dual Yang-Mills gauge fields~\cite{OV2}.
The exact
classical effective action for the gravitational, antisymmetric tensor
and gauge degrees of freedom
was   given in~\cite{KM3} and   derived
independently in~\cite{H1} using sigma-model techniques (see~\cite{M1,M4}
for reviews). In ref.~\cite{AH3},
this action was simplified using an auxiliary metric and shown to be
Weyl invariant at the classical level in four dimensions. A
dual form of this action was found; in four
dimensions, the dual geometry is self-dual gravity without torsion coupled to
a scalar field.

        The heterotic sigma-models which describe the target spaces of (2,1)
strings have been discussed in~\cite{HW,H3,H2}.
The
geometry is Hermitean with torsion and the field equations imply that the
curvature with torsion is self-dual in four dimensions, or satisifies
generalised
self-duality equations in higher dimensions. The conditions
under which these models have
isometry symmetries were analysed in ref.~\cite{AH1}, while the gauging of
such isometries and the construction of manifestly (2,1) supersymmetric
gauged actions were discussed in~\cite{AH1,AH2,MAth,KKL}.

This paper is organised as follows. In section~\ref{sect2}, we
discuss untwisted and twisted $(p,q)$ supersymmetry in two dimensions
and introduce a superspace for the general $(p,q)$ superalgebra. In
section~\ref{sect3} we construct the corresponding two dimensional
non-linear sigma models on target spaces of general signature, and
derive the geometric conditions imposed by supersymmetry.  In
section~\ref{sect:extsup}, we give a superspace formulation of the models
with twisted $(p,q)$ supersymmetry and discuss their isometry
symmetries. In section~\ref{sect:u21}, we review the geometry and the
extended superspace formulation of the sigma model with the usual
(2,1) supersymmetry. An extended superspace formulation of the sigma model with
twisted (2,1) supersymmetry is given in section~\ref{sect:t21}. In
section~\ref{sect:22} we discuss the various possible $N=2$ sigma
models and in particular give superspace formulations of the
pseudo-K\"{a}hler sigma models with or without torsion. We summarise
the results in section~\ref{sect:twistcomm}, and close with some remarks on a
reformulation
with \lq double numbers'.

\section{Superalgebras and Superspaces}

\label{sect2}

In two-dimensional Minkowski space, the   global supersymmetry algebra
of type $(p,q)$ was defined in
ref.~\cite{HW}. There also exists a twisted form of this algebra~\cite{H1} and
the general case is
\begin{eqnarray}
\left\{ Q^{I}_{+},Q^{J}_{+} \right\} & = & 2\eta^{IJ} P_+ , \ \
\ \ \left\{ Q^{I'}_{-}, Q^{J'}_{-} \right\} = 2\eta^{I' J'}P_- ,\nonumber \\
\left\{ Q^{I}_{+}, Q^{J'}_{-} \right\} & = & 0 ,
\label{susy}
\end{eqnarray}
where $Q_{+}^{I}$, $I=1, \ldots ,p$, are the $p$ positive-chirality
supersymmetry
charges, $Q^{I'}_{-}$, $I' = 1, \ldots ,q$ are the $q$ negative-chirality
charges and $+, -$ are chiral spinor indices; our   superspace
conventions are as in~\cite{GHR}.
The supercharges $Q_\pm$ are 1-component Majorana-Weyl spinors which, in our
conventions, are real,
$Q_\pm ^* =Q_\pm$.
Consider the right-handed superalgebra generated by the $Q_+^I$.
In the conventional (untwisted) superalgebra of ~\cite{HW},
$\eta^{IJ}=\delta^{IJ}$, while in the general case
 $\eta^{IJ}$   in~(\ref{susy}) can be
an arbitrary symmetric matrix. If invertible, it can be brought to the form
\begin{equation}
\eta^{IJ} = \left( \begin{array}{cc} 1_{u} & 0 \\ 0 & -1_{t} \end{array}
\right)
\label{eta}
\end{equation}
with $u+t=p$.
Then $Q^I_+$ for $I=1,\dots, u$ are normal supersymmetries that square to
$P_+$, while
 $Q^I_+$ for $I=u+1,\dots, p$ are twisted supersymmetries that square to
$-P_+$, and
we refer to the superalgebra as being twisted.
This can be generalised further to allow non-invertible metrics
\begin{equation}
\eta^{IJ} = \left( \begin{array}{ccc} 1_{u} & 0 &0 \\ 0 & -1_{t} &0
\\ 0 &0 &0_v
\end{array} \right)
\label{etaxx}
\end{equation}
with $v$ zeroes as well as $u$ $+1$'s and $t$ $-1$'s ($t+u+v=p$); there would
then be
$v$ nilpotent supercharges $Q^I_+$ for $I=u+t+1,\dots, p$ (i.~e.\ $Q^2=0$).
Note that for e.g. the twisted (2,0) algebra, the supercharges
$Q^\pm_+=Q^1_+\pm Q^2_+$ are each nilpotent, $(Q^\pm)^2=0$, but they do not
anti-commute with each other.
It would be interesting to study the cohomology associated with such nilpotent
supercharges.
The discussion of the left-handed superalgebra generated by the $Q_-^I$ is
similar, and there are corresponding expressions
for $\eta^{I' J'}$ with $p$
and $q$ interchanged.

The above can be extended further to allow central charges
$Z^{IJ'}$ with
\begin{equation}
\left\{ Q^{I}_{+}, Q^{J'}_{-} \right\}   =   Z^{IJ'} ,
\end{equation}
or vectorial charges $X_+^{IJ},X_-^{I'J'}$ with
\be
\left\{ Q^{I}_{+},Q^{J}_{+} \right\}   =   X_+^{IJ} , \ \
\ \ \left\{ Q^{I'}_{-}, Q^{J'}_{-} \right\} = X_-^{I'J'},
\label{susydsd}
\ee
but this will not be discussed further here.

Twisted superalgebras are possible in higher dimensions also; for instance the
ten-dimensional type II* string theories related by timelike T-duality
to the usual type II superstring theories
have twisted IIA or IIB superalgebras in 10 dimensions~\cite{CMHnew}.

It is straightforward to introduce a
superspace for the general $(p,q)$ superalgebra~(\ref{susy}).
There are two real bosonic coordinates $\sigma^{+}=\sigma^1+\sigma^2$,
$\sigma^{-}=\sigma^1-\sigma^2$, $p$ real positive-chirality Fermi coordinates
$\theta_{I}^{+}$ and
$q$   real negative-chirality Fermi coordinates $\theta^{-}_{I'}$. The
supersymmetry generators
\begin{equation}
Q_{+}^{I} = \frac{\partial}{\partial \theta_{I}^{+}} -i\eta^{IJ} \theta_{J}^{+}
\frac{\partial}{\partial \sigma^{+}}  , \ \ \ \ Q_{-}^{I'} =
\frac{\partial}{\partial \theta_{I'}^{-}} - i\eta^{I' J' } \theta^{-}_{J'}
\frac{\partial}{\partial \sigma^{+}} ,
\label{Qs}
\end{equation}
satisfy the superalgebra~(\ref{susy}); the corresponding supercovariant
derivatives
are
\begin{equation}
D_{+}^{I} = \frac{\partial}{\partial \theta_{I}^{+}} +i\eta^{IJ} \theta_{J}^{+}
\frac{\partial}{\partial \sigma^{+}}  , \ \ \ \ D_{-}^{I'} =
\frac{\partial}{\partial \theta_{I'}^{-}} + i\eta^{I' J' } \theta^{-}_{J'}
\frac{\partial}{\partial \sigma^{+}} ,
\label{Ds}
\end{equation}
and satisfy the anticommutators
\begin{eqnarray}
\left\{ D_{+}^{I} , D_{+}^{J} \right\} & = & 2i\eta^{IJ} \partial_{+} ,
\ \ \left\{ D_{-}^{I'},D_{-}^{J'} \right\} = 2i\eta^{I' J'}\partial_{-} ,
\nonumber \\ \left\{ D_{+}^{I}, D_{-}^{I'} \right\} & = & 0 .
\label{antiDs}
\end{eqnarray}

For Minkowski   world-sheets, there are one-component Majorana-Weyl spinors,
but for
Euclidean signature
there are no Majorana-Weyl spinors, so the analysis is different.
A Dirac spinor
\begin{equation}
\psi_a= \left( \begin{array}{cc} \psi_+ \\ \psi_- \end{array} \right)
\label{dira}
\end{equation}
has two complex components $\psi_\pm$. One can impose a Majorana condition
$(\psi_+)^*=\psi_-$ or a pseudo-Majorana condition
$(\psi_+)^*=-\psi_-$, or a Weyl condition  $\psi_+=0$ or $\psi_-=0$; there are
thus various types
of minimal spinor with 2 real components, but none with 1 component.

There are then various types of superalgebras in two Euclidean dimensions.
There   is a $(p,q)$ algebra with $p$ right-handed Weyl supercharges with
complex
components
$Q_+^I$ and $q$ left-handed Weyl supercharges with complex components
$Q_-^{I'}$, and the superalgebra is again~(\ref{susy}), but with all charges
complex, and
$P_\pm \equiv P_1
\pm i P_2$.
For $N$ Majorana spinors $Q_a^I$, $I=1,\dots , N$, the general algebra (without
central charges or
extra vector charges) is
\be
\left\{ Q^{I}_{a},Q^{J}_{b} \right\}   =   M^{IJ} P_\mu (\gamma ^\mu C)_{ab}
+N^{IJ} P_\mu (\gamma ^3\gamma ^\mu C)_{ab}
\ee
where $C$ is the two-dimensional charge conjugation matrix, the $\gamma^\mu$
are two-dimensional
Dirac matrices, $\gamma^3 = i\gamma^0 \gamma^1$ and $M^{IJ}$, $N^{IJ}$ are some
symmetric
matrices. The matrix $M^{IJ}$ can be taken to be diagonal with eigenvalues
$+1,-1 $ and $0$, as in
(\ref{etaxx}).  This can be obtained from the $(N,N)$ algebra with $N$
left-handed and $N$
right-handed Weyl supercharges
by imposing the Majorana condition $(Q^I_+)^*=Q_-^I$, with
$M^{IJ}={1\over 2} (\eta^{IJ}+\eta ^{I'J'})$ and
$N^{IJ}={1\over 2}(\eta^{IJ}-\eta ^{I'J'})$.
For pseudo-Majorana supercharges, the result is similar.
The general $(N,M,r,s)$ superalgebra with $N$ Majorana supercharges, $M$
pseudo-Majorana
supercharges,
$r$ right-handed Weyl supercharges and
$s$ left-handed Weyl supercharges can be obtained from the
$(p,q)$ superalgebra with $p=N+M+r$, $q=N+M+s$, by imposing the Majorana
condition
$(Q^I_+)^*=Q_-^{I'}$ for $I=I'=1,...,N$
and
the pseudo-Majorana  condition
$(Q^I_+)^*=-Q_-^{I'}$ for $I=I'=N+1,...,M+N$.
Thus all cases are contained in the Euclidean $(p,q)$ algebra, and much of the
analysis of the
Minkwoski $(p,q)$ models carries over to the Euclidean $(p,q)$ theories; in
particular, the
Eulidean $(p,q)$ superspace
has   two complex bosonic coordinates $\sigma^{\pm}=\sigma^1\pm i\sigma^2$,
  $p$ complex positive-chirality Fermi coordinates $\theta_{I}^{+}$ and
$q$   complex negative-chirality ones $\theta^{-}_{I'}$, with supercharges and
derivatives again given
by~(\ref{Qs}) and~(\ref{Ds}).

\section{ $(p,q)$ Sigma Models with General Target Space Signature}

\label{sect3}

We now turn to the  construction of non-linear
two dimensional sigma models with twisted or
untwisted ($p,q$) supersymmetry on target spaces of arbitrary signature.

It is convenient to first consider the
(1,1) supersymmetric sigma model with superspace action~\cite{GHR}
\begin{equation}
S_{(1,1)} = \int d^2 \sigma d\theta^+ d\theta^- [ g_{ij}(\phi ) + b_{ij} (\phi
)]
D_+ \phi^i D_- \phi^j  ,
\label{S11}
\end{equation}
where the $\phi^i$ are superfields which can be
viewed as coordinates on some $D$-dimensional
manifold $M$ with metric $g_{ij}$ and torsion 3-form $H$ given by the curl of
the
antisymmetric tensor $b_{ij}$,
\begin{equation}
H_{ijk} = \frac{3}{2}\partial_{[i}b_{jk]} .
\label{H}
\end{equation}
The action~(\ref{S11}) is invariant under (1,1) supersymmetry, general
coordinate transformations on the target manifold $M$ and antisymmetric tensor
gauge transformations
\begin{equation}
\delta b_{ij} = \partial_{[i} \lambda_{j]} .
\label{atgt}
\end{equation}

This model will be conformally invariant at one-loop if there is a function
$\Phi$ such that
\begin{equation}
R^{(+)}_{ij} -\nabla_{(i}\nabla_{j)} \Phi -H^k{}_{ij} \nabla_k \Phi =0,
\label{oneloop}
\end{equation}
where $R^{(+)}_{ij}$ is the Ricci tensor for a connection with torsion. We
define the connections with torsion
\begin{equation}
\Gamma^{(\pm )i}_{jk} = \left\{ \begin{array}{c} i \\ jk \end{array}
\right\} \pm H^i{}_{jk}
\label{gamm}
\end{equation}
where $ \left\{ \begin{array}{c} i \\ jk \end{array} \right\}$ is the
Christoffel connection, and
the corresponding covariant derivatives $\nabla^{(\pm )}$. The curvature
and Ricci tensors with torsion are
\begin{equation}
R^{(+)k}_{lij} = \partial_i \Gamma^{(+)k}_{jl} -\partial_j \Gamma^{(+)k}_{il}
+\Gamma^{(+)k}_{im}\Gamma^{(+)m}_{jl} -\Gamma^{(+)k}_{jm}\Gamma^{(+)m}_{il}
,
\ \ R^{(+)}_{ij} = R^{(+)k}_{ikj} .
\label{RimRic}
\end{equation}
The equation~(\ref{oneloop}) can be obtained by varying the action
\begin{equation}
S=\int d^D x e^{-2\Phi} \sqrt{|g|} \left( R-\frac{1}{3}H^2 +4 (\nabla \Phi )^2
\right) .
\end{equation}

We now seek the conditions on the target space geometry under which
the (1,1)
superspace action~(\ref{S11})  is invariant under extra supersymmetries,
generalising
the analysis of~\cite{HW,H3,H2,H02} to arbitrary signature and giving a
superspace
derivation of the results
of~\cite{H1}. If there are
$p-1$ right-handed and $q-1$ left-handed extra supersymmetry transformations,
then
they must be of
the form
\begin{equation}
\delta \phi^i =\varepsilon^r T^i_{(+)rj} D_+ \phi^j +\varepsilon^{r'}
T^{i}_{(-)r'j}
D_- \phi^j
\label{anysusy}
\end{equation}
 for some tensors  $(T_{(+)r})^i{}_j$,
$(T_{(-)r'})^i{}_j$
with $r=1,\ldots ,p-1$ and  $r' = 1,\ldots ,q-1$.
Invariance of the action~(\ref{S11}) requires that
  the tensors $T^i_{(+)rj}$,
$T^i_{(-)r'j}$ satisfy
\begin{eqnarray}
g_{ki}T^k_{(+)rj} + g_{kj} T^{k}_{(+)ri} & = & 0, \nonumber \\ g_{ki}
T^{k}_{(-)r'j} +g_{kj} T^{k}_{(-)r'i} & = & 0  ,
\label{Tcovconst}
\end{eqnarray}
and
\begin{equation}
\nabla_{k}^{(+)} T^{i}_{(+)rj} =\nabla^{(-)}_k T^{i}_{(-)r'j} =0 .
\label{Tcovconst2}
\end{equation}
If the supersymmetry transformations~(\ref{anysusy}) are to satisfy a
superalgebra, which
may be twisted or
untwisted, then the
matrices
$T_{r(+)}$ and $T_{r'(-)}$
must satisfy
  anticommutation relations of the form
\be
\left\{ T_{(+)r},T_{(+)s}\right\}   =   -2\eta_{rs} , \ \
\ \ \left\{ T_{(-)r'} , T_{(-)s'}  \right\} = -2\eta_{r' s'},
\label{sugfsydsd}
\ee
for some metrics $\eta^{rs}$, $\eta^{r' s'}$.
In addition, the generalised  Nijenhuis concomitants  ${\cal N}
(T^{r}_{+},T^{s}_{+} )$
and ${\cal N} (T^{r'}_{-}, T^{s'}_{-})$ must vanish.
For any  (1,1) tensors $T_1$ and $T_2$ the   generalised  Nijenhuis concomitant
is defined
by~\cite{FN}
\begin{eqnarray}
{\cal N} (T_1 ,T_2 )^i{}_{jk} & = & {T_1}^l{}_{j}\partial_l {T_2}^i{}_k
-{T_1}^l{}_{k}\partial_l {T_2}^i{}_j -{T_1}^i{}_l \partial_j {T_2}^l{}_k -
{T_1}^i{}_l \partial_k {T_2}^l{}_j \nonumber \\ & & +(1\to 2 )
{}
\end{eqnarray}
so that ${\cal N}(T_1 ,T_2 ) = {\cal N} (T_2, T_1)$
and ${\cal N} (T_1 ,T_2 )^i{}_{jk}$ is antisymmetric in the indices $j,k$. Then
${1\over 4}{\cal N}(T, T)
\equiv {\cal N} (T)$ is the usual Nijenhuis tensor of $T$,
\begin{equation}
{\cal N}^k{}_{ij} (T) = T^l{}_{i} T^k{}_{[j,l]} -T^l{}_j T^k{}_{[i,l]}
{}.
\label{Nijen}
\ee
The condition $ {\cal N} (T)=0$ implies that $T$ is integrable, i.e. that a
coordinate system can
be chosen in which it is constant.
However, if there are several integrable such tensors, it will usually not be
possible
to choose
coordinates in which they are simultaneously integrable.

If the above conditions are satisfied,  then the supersymmetry
transformations~(\ref{anysusy})
together with the
manifest (1,1) supersymmetries satisfy the algebra~(\ref{susy}) with
\begin{equation}
\eta^{IJ} = \left( \begin{array}{cc} 1 & 0 \\ 0 & \eta^{rs} \end{array}
\right) , \ \ \ \ \eta^{I'J'} = \left( \begin{array}{cc} 1 & 0 \\ 0 &
\eta^{r's'} \end{array}
\right) .
\label{etars}
\end{equation}
Diagonalising
$\eta^{rs}$, $\eta^{r' s'}$, we find that each tensor $T$ squares to either
$+1$, $-1$  or $0$; those
satisfying $T^2=-1$ are complex structures while those satisfying $T^2=1$
are sometimes referred to as real structures (as in~\cite{NJH,BGPPR}) and
sometimes as almost
product structures
(as in~\cite{GHR}).

Consider first the case of the right-handed supersymmetries with the tensors
$T_r=T_{(+)r}$.
Each is Hermitean, $T_{ij}=-T_{ji}$, and covariantly constant with respect to
the
connection $ \Gamma^{(+)}$,
and so the $p-1$ tensors $T_r$ must be singlets under
the holonomy group ${\cal H} $ of $ \Gamma^{(+)}$. We will restrict ourselves
to the cases in
which the holonomy is
irreducible.
 For  signature $(m,n)$, ${\cal H} $ is $O(n,m)$, or a subgroup thereof, as
the metric
with  signature $(m,n)$ is covariantly constant.
There  will be a covariantly constant complex structure
$J$, with $J^2=-1$,
if $m,n$ are even, $n=2n_1,m=2n_2$, so that the signature is
$(2n_1,2n_2)$, and if ${\cal H} \subseteq U(n_1,n_2)$.
If there are two covariantly constant complex structures, $I,J$, then $K=IJ$ is
a third
covariantly constant complex structure and the $I,J,K$ satisfy the  quaternion
algebra
\begin{eqnarray}
& I^2=J^2=K^2=- {1}, \nonumber \\
&IJ=-JI=K, \ \ JK=-KJ=I, \ \  KI=-IK=J
\label{quat}
\end{eqnarray}
with $I,J,K$ satisfying $SO(3)$ commutation relations.
This requires that the holonomy group is contained  in $USp(2m)$ for Euclidean
spaces of even complex dimension $n=2m$ (where $USp(2m)$ is compact, with the
convention that $USp(2)=SU(2)$; we use the definitions of groups and their
non-compact forms  given in~\cite{Gilmore}).
For spaces of signature $(4n,4m)$, this requires that the holonomy is contained
in
$USp(2n,2m)$ (this is the subgroup of $U(2n,2m)$ preserving a symplectic
structure).

For a real structure $S$ satisfying $S^2=1$, the hermiticity condition implies
that the metric, if
it is to be non-degenerate, has to be of signature $(m,m)$, and the holonomy
group has to be in
$GL(m,\R)$.
If there are two real structures, $S,T$ with $\{ S,T \} =0$, then $J=ST$ is a
complex structure and
$J,S,T$ must
satisfy the pseudo-quaternion algebra
\begin{eqnarray}
& J^2 =- {1}, \ \ S^2=T^2={1}, \nonumber \\
&ST=-TS=-J, \ \ TJ=-JT=S,  \ \ JS=-SJ=T
\label{psquat}
\end{eqnarray}
with $J,S,T$ satisfying $SO(2,1)$ commutation relations, so that there is a
pseudo-quaternionic
structure ~\cite{NJH,BGPPR}. Similarly, if there is a complex structure $J$ and
a real structure $S$ with $\{ S,J \} =0$,
then $T=JS$ is another real structure and  $J,S,T$ again
satisfy the pseudo-quaternion algebra~(\ref{psquat}).
The existence of such a covariantly constant pseudo-quaternionic structure
requires that $m$ is
even, $m=2k$, and the holonomy is in $Sp(2k,\R)$.
If $p>4$, the tensors $T$ satisfy an octonion or pseudo-octonion algebra and
the holonomy must be
trivial. Similar results apply for the left-handed supersymmetries, the number
of which depends on
the holonomy of the connection $\Gamma ^{(-)}$.

 The   currents
\begin{equation}
j_{(\pm)r}  = \frac{1}{2} T^{ij}_{(\pm)r} \psi_i \psi_j
\end{equation}
generate left and right handed Ka\v{c}-Moody algebras.
The right-handed currents $j_{(+)r}$ generate an affine $SO(2) $ or $SO(3)$ if
there are
$p=2$ or  $p=4$
untwisted supersymmetries,  and
an affine $SO(1,1)$  if  $p=2$
 and one of the supersymmetries is twisted, and
an affine   $SO(2,1)$ if  $p=4$
 and two of the supersymmetries are twisted.
In the latter case,
the $SO(2,1)$ Ka\v{c}-Moody algebra   is part of a non-compact
twisted form of the (small) $N=4$ superconformal algebra with global limit
given
by~(\ref{susy}), where $\eta^{IJ}$ is the $O(2,2)$ invariant
metric~\cite{H1}.

In the special case in which the torsion vanishes, then
$\Gamma^{(+)}=\Gamma^{(-)}=\Gamma$
and the number of left-handed supersymmetries is the same as the number of
right-handed supersymmetries, $p=q$.
For $(2,2)$ untwisted supersymmetry  the geometry is K\"{a}hler, for $(4,4)$
untwisted supersymmetry
the geometry is hyper-K\"{a}hler, while for $(2,2)$  twisted supersymmetry we
shall
call the geometry
pseudo-K\"{a}hler, and for
$(4,4)$  twisted supersymmetry  we shall call   the geometry
pseudo-hyper-K\"{a}hler.
The pseudo-K\"{a}hler geometry shares many of the features of K\"{a}hler
geometry; in
particular, the metric
can in both cases be given in terms of a scalar potential, as we shall see in
section~\ref{sect:t21}.

\section{Extended Superspace and Isometries}

\label{sect:extsup}

  A superspace formulation of the models with twisted  $(p,q)$ supersymmetry
can be given in $(p,q)$
superspace  using a formalism which generalises that proposed by Howe and
Papadopoulos
in~\cite{HP1,HP2}. Let
\begin{equation}
(\sigma^{\pm},\theta_{\mu}^{-} ,\theta_{\mu '}^{+},\tilde{\theta}_{\tnu}^{-}
,\ttheta_{\tnu
'}^{+})
\nonumber
\end{equation}
with
$\mu =0,
\ldots ,u$, $\tmu =u+1 ,\dots ,p-1$ and $\nu ' = 0, \ldots , v$, $\tnu '= v+1,
\ldots ,q-1$ ($1\leq u\leq p-1$, $1\leq v\leq q-1$) be the superspace
coordinates. The
nonvanishing anticommutators
of the flat superspace derivatives
$D_{\mu +}$ and $D_{\mu '-}$ are
\begin{eqnarray}
\{ D_{\mu +},D_{\nu +} \} & = & 2i \delta_{\mu \nu} \partial_{+} , \ \ \ \
\{ D_{\mu '-}, D_{\nu '-} \} =2i\delta_{\mu ' \nu '} \partial_- , \nonumber \\
\{ D_{\tmu  +},D_{\tnu  +} \} & = & -2i \delta_{\tmu \tnu} \partial_{+} , \ \
\ \
\{ D_{\tmu '-}, D_{\tnu '-} \} =-2i\delta_{\tmu ' \tnu '} \partial_- .
\end{eqnarray}
$D_{\mu +}$ and $D_{\mu '-}$ anticommute with the supercharges
$Q_{\mu +}$ and $Q_{\mu '-}$, while $D_{\tmu +}$ and $D_{\tmu '-}$ anticommute
with $Q_{\tmu +}$ and $Q_{\tmu '-}$. The generalised $(p,q)$ non-linear sigma
model is described by a
superfield $\varphi^i$ which is a map from the $(p,q)$ superspace to $M$. The
chirality
constraints~\cite{HP1,HP2}
\begin{eqnarray}
D_{r+} \varphi^i & = & T^i_{(+)rj} D_{0+} \varphi^j , \ \ \ \ r=1,\ldots ,p-1 ,
\nonumber \\ D_{r'-}\varphi^i & = & T^i_{(-)r'j} D_{0-} \varphi^j , \ \ \ \
r' = 1,\ldots ,q-1 ,
\label{pqconstr}
\end{eqnarray}
imply that the $(p,q)$ supersymmetry transformation of either type generated
by ($Q_{\mu +}$, $Q_{\mu '-}$)  and ($Q_{\tmu +}$, $Q_{\tmu '-}$) reduce to
the
transformations~(\ref{anysusy})
on expanding into (1,1) superfields.

The twisted or untwisted $(p,q)$-supersymmetric sigma model action in
the corresponding $(p,q)$
superspace is then~\cite{HP1,HP2}
\begin{equation}
S= -i \left[ \int d^2 \sigma d\theta_0^+ d\theta_0^- g_{ij} D_{0+}
\varphi^i D_{0-}\varphi^j + \int d^2 \sigma dt d\theta_0^+ d\theta_0^-
H_{ijk} \partial_t  D_{0+}\varphi^j D_{0-}\varphi^k \right] ,
\label{Spq}
\end{equation}
where the $(p,q)$ superfields satisfy the constraints~(\ref{pqconstr}). If
eqs.~(\ref{Tcovconst}) and~(\ref{Tcovconst2}) hold, then using the
constraints~(\ref{pqconstr}) it can be shown that the action~(\ref{Spq})
is independent of the extra supercoordinates ($\theta_r$,
$\theta_{r'})$, and
as a result is invariant (up to surface terms) under the non-manifest
supersymmetries generated by ($Q_{r+}$, $Q_{r '-}$) and ($Q_{\tildar +}$,
$Q_{\tildar '-}$).

Now consider $(p,q)$ infinitesimal superspace transformations of the form
\begin{equation}
\delta \varphi^i = \lambda^a \xi^i_a (\varphi )
\label{rigidtransf}
\end{equation}
with constant parameters $\lambda^a$. These will constitute proper symmetries
of the sigma model
action~(\ref{Spq}) if the metric and torsion are Lie invariant,
\begin{equation}
({\cal L}_a g)_{ij} = 0, \ \ \ \ ({\cal L}_a H)_{ijk} = 0 ,
\end{equation}
and if in addition
\begin{equation}
{\cal L}_a T_{(+)r} = {\cal L}_a T_{(-)r'} = 0,
\end{equation}
i.~e.\ the real or complex structures are also Lie invariant. Then the
$\xi^i_a$ are Killing vectors which are
holomorphic with respect to each complex structure, or \lq holomorphic' in a
generalised sense with
respect to each real structure. This implies locally on $M$ that
\begin{equation}
\xi^{i}_{a} H_{ijk} = 2\partial_{[j}u_{k]a} ,
\end{equation}
where $u$ is a locally defined one-form $u_{ia}$ which is determined in every
coordinate patch of $M$
up to an exact Lie-algebra valued one-form. It follows that there are
generalised Killing potentials
$X_{(+)ra}$, $X_{(-)r'a}$ satisfying
\begin{equation}
g_{ij}\xi^j_a +u_{ia} = T^{j}_{(+)ri} \partial_j X_{(+)ra} = T^{j}_{(-)r'i}
\partial_j X_{(-)r'a}
\end{equation}
for every $r=1,\ldots ,p-1$ and $r' =1,\ldots ,q-1$.

\section{(2,1) Sigma Models }

\label{sect:u21}

        In this section we review the (2,1) sigma model with untwisted
supersymmetry; the model with twisted (2,1) supersymmetry will be discussed
in section~\ref{sect:t21}. The
geometric conditions for the (1,1) model to have
untwisted (2,1) world-sheet supersymmetry were first obtained in
refs.~\cite{HW}, and follow from the general discussion given above. The
manifold must be
complex (with dimension $D=2n$) with metric
$g_{ij}$ of signature $(2m_1,2m_2)$ with $m_1+m_2=n$ and a complex structure
$J^i{}_{j}$ which
is covariantly constant with respect to the connection with torsion
$\Gamma^{(+)}$ defined in~(\ref{gamm})
and with respect to which the metric is Hermitean, so that $J_{ij} = g_{ik}
J^{k}{}_{j}$ is antisymmetric. Introducing complex coordinates $z^\alpha$,
$\overline{z}^{\overline{\beta}}=(z^{\beta} )^{*}$ in which the complex
structure is constant and diagonal,
\begin{equation}
J^i{}_{j} = i \left( \begin{array}{cc} \delta^{\beta}_{\alpha} & 0 \\ 0 &
-\delta^{\overline{\beta}}_{\overline{\alpha}} \end{array} \right) ,
\end{equation}
any $N$-form
can be decomposed into a set of $(r,s)$ forms with $r$ factors of $dz$
and $s$ factors of $d\overline{z}$, where $r+s =N$. The conditions above then
imply that the (0,3) and (3,0) parts  of the three-form $H$
vanish and $H$ is given in terms of the fundamental two-form
\begin{equation}
J = \frac{1}{2} J_{ij}d\phi^i \wedge d\phi^j = -ig_{\alpha \overline{\beta}}
dz^\alpha \wedge \overline{z}^{\overline{\beta}}
\label{fundJ}
\end{equation}
by
\begin{equation}
H=i\left( \partial -\overline{\partial} \right) J .
\end{equation}
The exterior derivative decomposes in the complex coordinate system as
$d=\partial + \overline{\partial}$, so the closure of the three-form $H$
implies
\begin{equation}
i\partial \overline{\partial} J = 0 .
\end{equation}
It follows that locally a (1,0) form $k=k_\alpha dz^\alpha$ exists such that
\begin{equation}
J = i\left( \partial \kbar + \overline{\partial} k \right) .
\end{equation}
The metric and torsion potential are then given (in a suitable gauge) by
\begin{eqnarray}
g_{\alpha \bbar} & = & \partial_\alpha \overline{k}_{\overline{\beta}} +
\partial_{\overline{\beta}}k_{\alpha} \nonumber \\ b_{\alpha \overline{\beta}}
& = & \partial_\alpha \kbar_{\bbar} -\partial_{\overline{\beta}} k_\alpha .
\label{complgeom}
\end{eqnarray}
If $k_\alpha = \partial_\alpha K$ for some $K$ then the torsion vanishes and
the manifold is K\"{a}hler
with K\"{a}hler potential $K$ and the (2,1) supersymmetric
model in fact has (2,2) supersymmetry, but if $dk \neq 0$, then $M$ is a
Hermitean manifold with torsion~\cite{HW}. The metric and torsion are
invariant under~\cite{AH1}
\begin{equation}
\delta k_\alpha = i\partial_\alpha \chi +\theta_\alpha
\label{ksymm}
\end{equation}
where $\chi$ is real and $\theta_\alpha $ is holomorphic,
$\partial_{\overline{\beta}}\theta_\alpha = 0$, but $b_{\alpha \bbar}$
as defined in~(\ref{complgeom}) transforms as
\begin{equation}
\delta b_{\alpha \bbar} = -2i\partial_\alpha \partial_{\bbar} \chi ,
\end{equation}
which is an antisymmetric gauge transformation~(\ref{atgt}) with parameter
$\lambda_\alpha =
2i\partial_\alpha \chi$.

Much of the above structure can be found using superspace methods.
We start by seeking the  most general (2,1) supersymmetric sigma-model
 that can be written in  a (2,1) superspace parametrised by $\sigma^\mu$,
$\theta^+$, $\thetbar^{+}$, $\theta^-$, where $\theta^+ = \theta_1 +i\theta_2$
is a complex Weyl spinor and the corresponding supercovariant derivatives are
\begin{eqnarray}
D_+ & = & \frac{\partial}{\partial \theta^+} +i\thetbar^+ \partial_+  , \ \ \ \
\Dbar_+ = \frac{\partial}{\partial \thetbar^+} +i\theta^+ \partial_+
,\nonumber \\ D_- & = & \frac{\partial}{\partial \theta^-} +i\theta^-
\partial_- ,
\end{eqnarray}
so that
\begin{equation}
\{ D_+ ,\Dbar_+ \} =2i\partial_+  , \ \ \ \ \{D_+ ,D_- \} = \{ \Dbar_+ ,D_- \}
=0
{}.
\end{equation}
We introduce complex (2,1) scalar  superfields $ \varphi^\alpha$, $
\overline{\varphi}^{\abar}=(\varphi^\alpha)^*$ satisfying the chiral constraint
\begin{equation}
\overline{D}_{+}\varphi^\alpha = 0  ,\ \ \ \ D_+ \overline{\varphi}^{\abar} =
0.
\end{equation}
The lowest components $\varphi^\alpha |_{\theta =0} =z^\alpha $ of the
superfields
are   bosonic complex coordinates of the target space. The general sigma-model
action
is~\cite{DS}
\begin{equation}
S=i\int d^2 \sigma d\theta^{+}\overline{\theta}^{+}d\theta^{-} \left(
k_\alpha D_- \varphi^\alpha -\overline{k}_{\overline{\alpha}} D_-
\overline{\varphi}^{\overline{\alpha}}\right)
\label{I21real}
\end{equation}
for some local vector potentials $k_\alpha ( \varphi^\alpha,
\overline{\varphi}^{\abar})$,   $\overline{k}_{\overline{\alpha}} (
\varphi^\alpha,
\overline{\varphi}^{\abar})$, which are required to be complex conjugate if the
action~(\ref{I21real}) is to be
real,
$\overline{k}_{\overline{\alpha}} =(k_\alpha)^*$.
Expanding in components, the bosonic part of the action is
a bosonic sigma-model with metric $g_{\alpha \bbar}$ and torsion potential
$b_{\alpha \bbar}$
given in terms of $k$ by~(\ref{complgeom}), so that we find the geometry
described above.
In particular, if  $k_\alpha=\partial _\alpha K $ for some scalar $K$, then the
torsion vanishes and the
metric is given by
\begin{equation}
g_{\alpha \bbar} =\partial_\alpha \partial_{\bbar} K
\label{kahlermetric}
\end{equation}
so it is K\"{a}hler.

The additional geometric conditions under which the model has isometry
symmetries have been analysed in ref.~\cite{AH1}. There it was shown
that the geometry determines the potentials $\chi$ and $\theta$ that appear
in eq.~(\ref{ksymm}). The construction
of gauged (2,1) superspace actions was discussed in refs.~\cite{AH1,AH2,MAth}.

 It will be useful to define the
vector
\begin{equation}
w^i = H_{jkl}J^{ij}J^{kl}
\label{w}
\end{equation}
together with the $U(1)$ part of the curvature
\begin{equation}
C_{ij}^{(+)} = J^l{}_{k} R^{(+) k}_{lij}
\label{U1R}
\end{equation}
and the $U(1)$ part of the connection~(\ref{gamm}),
\begin{equation}
\Gamma_{i}^{(+)} = J^k{}_{j} \Gamma^{(+)j}_{ik} = i\left(
\Gamma^{(+)\alpha}_{i\alpha}
-\Gamma^{(+)\overline{\alpha}}_{i\overline{\alpha}}
\right) .
\label{U1gamm}
\end{equation}
Note that $C_{ij}$ is a representative of the first Chern class, and that it
can be written as $C^{(+)}_{ij}=2\partial_{[i}\Gamma^{(+)}_{j]}$ in a
complex coordinate system. If the metric has Euclidean signature, the
holonomy of
any metric connection (including $\Gamma^{(\pm )}$) is contained
in $O(2n)$, while if it has signature $(2m_1 ,2m_2 )$ with
$m_1 +m_2 =n$, it will be contained in $O(2m_1 ,2m_2 )$.
As the complex structure is covariantly constant, the
holonomy ${\cal H} (\Gamma^{(+)})$ of the connection with
torsion $\Gamma^{(+)}$ is contained in $U(m_1 ,m_2 )$,
but it will be contained in $SU(m_1 ,m_2)$ if in addition
$C^{(+)}_{ij} =0$; a necessary condition for this is the vanishing of the
first Chern class.

It was shown in~\cite{Buscher,H2,H02} that geometries for which
\begin{equation}
\Gamma^{(+)}_i = 0
\label{Gam0}
\end{equation}
in some suitable choice of coordinate system will satisfy the conditions for
one-loop conformal invariance~(\ref{oneloop}) provided the dilaton is chosen as
\begin{equation}
\Phi =-\12 \log | \det g_{\alpha \bbar}| ,
\label{dilch}
\end{equation}
which implies
\begin{equation}
\partial_i \Phi =v_i .
\end{equation}
Moreover, the one-loop dilaton field equation is also satisfied for compact
manifolds, or for non-compact ones in which $\nabla \Phi$ falls off
sufficiently fast~\cite{H1}. This implies that ${\cal H} (\Gamma^{(+)})$
is contained in $SU(m_1 ,m_2 )$. These geometries generalise the
K\"{a}hler Ricci-flat or Calabi-Yau geometries, and reduce to these in
the special case in which $H=0$. However they are not the most general
solutions of the conditions~(\ref{oneloop})~\cite{H1}.

The condition that the connection $\Gamma^{(+)}$ has $SU(m_1 ,m_2 )$ holonomy
can be cast as a generalised self-duality condition on the curvature. Defining
the four-form
\begin{equation}
\phi^{ijkl} \equiv -3 J^{[ij}J^{kl]} ,
\end{equation}
the condition that ${\cal H} (\Gamma^{(+)}) \subseteq SU(m_1 ,m_2 )$ is
equivalent
to~\cite{AH3}
\begin{equation}
R^{(+)}_{ijkl} = \12 g_{im}g_{jn} \phi^{mnpq} R^{(+)}_{pqkl} .
\label{Rsd}
\end{equation}
For $D=4$, $\phi^{ijkl} = -\epsilon^{ijkl}$ and this is the usual
anti-self-duality condition.

The equation~(\ref{Gam0}) can be viewed as a field equation for the potential
$k_\alpha$, and can be obtained by varying the action~\cite{KM3,M1,H1}
\begin{equation}
S=\int d^D x \sqrt{|\det g_{\alpha \bbar}|},
\label{HKM}
\end{equation}
where $g_{\alpha \bbar}$ is given in terms of $k_\alpha$ by~(\ref{complgeom}).
This action can be rewritten as
\begin{equation}
S= \int d^D x |\det g_{ij} |^{1/4}
\label{HKMr}
\end{equation}
which is non-covariant but is invariant under volume-preserving
diffeomorphisms. This can be rewritten in the classically
equivalent alternative form~\cite{AH3}
\begin{equation}
S' = T'_4  \int d^D x |\gamma|^{1/4} \left[
\gamma^{ij}  g_{ij}    -(D-4)c \right]  ,\label{Hform4}
\end{equation}
where $\gamma_{ij}$ is an auxiliary metric, $\gamma =\det
\gamma_{ij}$ and $c,T'_4$ are (real)
constants. In the special case of four dimensions, the constant term
in the action~(\ref{Hform4}) vanishes and there is a generalised Weyl
symmetry under
\begin{equation}
\gamma_{ij} \rightarrow \omega (x) \gamma_{ij} .
\end{equation}

The dualisation of the action~(\ref{HKM})  was discussed in
ref.~\cite{AH3}. This is achieved by adding
a Lagrange multiplier term  imposing the constraint $g_{\alpha \bbar}
=\partial_\alpha \kbar_{\bbar} +\partial_{\bbar} k_\alpha$. The
vector potentials $k_\alpha ,k_{\abar}$ are then Lagrange multipliers for a
certain constraint, and solving this leads to a dual form
of the action~\cite{AH3}. In four dimensions, the dual geometry is self-dual
gravity without torsion coupled to a scalar field, while in $D>4$ dimensions
the dual geometry is Hermitean and determined by a $D-4$ form potential $K$
which generalises the K\"{a}hler potential of the four-dimensional case. The
coupling to the Yang-Mills fields is through a term $K\wedge tr (F\wedge F)$
and leads to a Uhlenbeck-Yau field equation $\tilde{J}^{ij}F_{ij}
=0$~\cite{AH3}.

\section{Twisted (2,1) Sigma Models}

\label{sect:t21}

        Consider now the case of space-time signature $(d,d)$, which was called
Kleinian in ref.~\cite{BGPPR}. We start by
considering the (2,1)
superspace formulation to obtain the geometry in a special coordinate system
(the analogue of the
complex coordinates of section~\ref{sect:u21}), then show how the same results
can be
obtained in a coordinate
independent manner using the results of section~\ref{sect3}.  Using the (2,1)
superspace introduced in section~\ref{sect2}, we
define
\begin{equation}
\theta _+ =
\theta^1_+ + \theta _+^2, \ \ \tilde \theta _+ =
\ttheta^1_+ - \ttheta _+^2
\end{equation}
and the supercharges
\begin{equation}
Q_+ = \frac{\partial}{\partial \theta_+} -\ttheta_+ \frac{\partial}{\partial
\sigma^+}  , \ \ \ \ \tilde{Q}_+ = \frac{\partial}{\partial \tilde{\theta}_+}
-\theta_+ \frac{\partial}{\partial \sigma^+},
\ee
satisfying the algebra
\begin{equation}
\{ Q_+ ,\tilde{Q}_+ \} = 2\partial_+  ,\ \ \ \ Q_+^2 = \tilde{Q}_+^2 =0 ,
\end{equation}
together with the superderivatives
\begin{equation}
D_+ = \frac{\partial}{\partial \theta^{+}} + \tilde{\theta}^{+}
\frac{\partial}{\partial \sigma^{+}} , \ \  \ \ \tilde{D}_+ =
\frac{\partial}{\partial \tilde{\theta}^{+}} + \theta^{+}
\frac{\partial}{\partial \sigma^{+}} ,
\label{twistsuper}
\end{equation}
which satisfy the anticommutators
\begin{eqnarray}
\left\{ D_+ ,D_+ \right\} & = & \left\{ \tilde{D}_+ ,\tilde{D}_+ \right\} = 0
\nonumber \\ \left\{ D_+ ,\tilde{D}_+ \right\} & = & 2\partial_{+} .
\end{eqnarray}
Thus the structure associated with $\theta _+,\tilde \theta _+ $ in the twisted
(2,1) case is
similar to that associated with $\theta _+,\bar \theta _+ $ in the untwisted
(2,1) case, with
the important difference that in the usual case $\theta _+,\bar \theta _+ $ are
complex and
related by $(\theta _+)^*=\bar \theta _+ $, while in the twisted case  $\theta
_+,\tilde \theta _+ $
are independent real coordinates.

        The twisted (2,1) supersymmetric sigma-model
can be
formulated in a twisted (2,1) extended superspace as follows. First we
introduce
chiral scalar superfields $U^\alpha$ and $\tilde{V}^{\tilde{\alpha}}$
satisfying
\begin{equation}
\tilde{D}_+ U^\alpha = 0 ,\ \  \ \ D_+ \tilde{V}^{\tilde{\alpha}} =0 .
\end{equation}
Note that as $\tilde{D}_+ , D_+$ are independent real derivatives, we take
$U^\alpha  ,
\tilde{V}^{\tilde{\alpha}}$ as independent real superfields.
Here $\alpha =1,....,n$ and $\tilde \alpha =1,....,\tilde n$ for some $n,\tilde
n$.
 The general twisted superspace action is
\begin{equation}
S = -\int d^2 \sigma d\theta^{+}d\tilde{\theta}^{+}d\theta^{-} \left(
k_\alpha D_- U^\alpha -\tilde{k}_{\tilde{\alpha}} D_-
\tilde{V}^{\tilde{\alpha}}
\right)
\label{Ie}
\end{equation}
for some independent real vector potentials
$k_\alpha (U^\alpha  ,
\tilde{V}^{\tilde{\alpha}}),\tilde{k}_{\tilde{\alpha}}(U^\alpha  ,
\tilde{V}^{\tilde{\alpha}})$.
The corresponding Lagrangian in (1,1) superspace can be obtained
by integrating over $\tilde{\theta}_{2}^{+}$. Up to a total derivative term, we
find
the action
\begin{equation}
S = \int d^2 \sigma d \tilde{\theta}_1^+ d\theta^- [g_{\alpha \tilde{\beta}}+
b_{\alpha \tilde{\beta}} ] D_{1+} u^\alpha D_- v^{\tilde{\beta}} ,
\end{equation}
 where $u,\tilde v$ are the lowest components of the superfields $U$ and
$\tilde{V}$.
 The metric and torsion potential are given by
\begin{eqnarray}
g_{\alpha \tilde{\beta}} & = & \partial_\alpha \tilde{k}_{\tilde{\beta}}
+\partial_{\tilde{\beta}}k_\alpha \label{realgeom1} 
\\ b_{\alpha \tilde{\beta}}
 & = & \partial_\alpha
\tilde{k}_{\tilde{\beta}} -\partial_{\tilde{\beta}} k_\alpha .
\label{realgeom2}
\end{eqnarray}
with $g_{\alpha {\beta}}=b_{\alpha {\beta}}=0$.
The target space line-element is
\begin{equation}
ds^2 = 2g_{\alpha \tilde{\beta}} (u,v) du^\alpha d\tilde{v}^{\tilde{\beta}}
\label{lineel}
\end{equation}
so that $\partial /\partial u^\alpha$ and $\partial / \partial
\tilde{v}^{\tilde{\beta}}$ are null vectors.
If $n\ne \tilde  n$, the metric constructed in this way is
  degenerate  in general; we will not consider
this case further
and restrict ourselves to the case $n= \tilde  n$.

The condition for the torsion to vanish is
\begin{equation}
k_\alpha = -\partial_\alpha \kappa  , \ \ \ \  \tilde{k}_{\tilde{\beta}}
=\partial_{\tilde{\beta}}\tilde{\kappa}
\label{condnoH}
\end{equation}
for some locally defined potentials $\kappa ,\tilde{\kappa}$. If this is
satisfied, then the metric is given in terms of a
scalar potential $\tilde K=\kappa -\tilde{\kappa}$,
\begin{equation}
g_{\alpha \tilde{\beta}} = \frac{\partial^2}{\partial u^\alpha \partial
v^{\tilde{\beta}}} \tilde{K},
\label{Kahl}
\end{equation}
giving a real-structure analogue of K\"{a}hler geometry, pseudo-K\"{a}hler
geometry.

The same geometry  can be obtained using the results of section~\ref{sect3} as
follows. The
(1,1) sigma models with  target space signature $(n,n)$
and a  covariantly constant real
structure $S$  will
have twisted (2,1) supersymmetry with global limit given by the supersymmetry
algebra~(\ref{susy}), where
$I,J=1,2$ and $\eta^{IJ} = \diag (1,-1)$. The holonomy is ${\cal H}
(\Gamma^{(+)}) \subseteq GL(n,\R)$. The integrable real structure $S$ squares
to $+1$,
\begin{equation}
S^i{}_{k}S^k{}_{j} = +\delta^i{}_j .
\end{equation}
Twisted (2,1) supersymmetry requires $S^i{}_{j}$ to be covariantly
constant with respect to the connection with torsion $\Gamma^{(+)}$,
\begin{equation}
\nabla^{(+)}_{k} S^i{}_{j} = 0
\label{Scovconst}
\end{equation}
and to be antisymmetric
\begin{equation}
S_{ij} = -S_{ji} .
\label{Sherm}
\end{equation}

        As  $S^i{}_j$ is integrable  (i.e. its Nijenhuis
tensor~(\ref{Nijen}) vanishes),   there is a coordinate system in which it is
constant and
diagonal. Choosing such adapted real coordinates $u^\alpha$,
$v^{\tilde{\alpha}}$
($\alpha =1,2$; $\tilde{\alpha} = 1,2$), the real structure takes the form
\begin{equation}
S^i{}_j = \left( \begin{array}{cc} \delta^{\alpha}_{\beta} & 0 \\ 0 &
-\delta^{\tilde{\alpha}}_{\tilde{\beta}} \end{array} \right) .
\end{equation}
The fundamental two-form
is then
\begin{equation}
S = \frac{1}{2} S_{ij}d\phi^i \wedge d\phi^j =  -g_{\alpha \tilde{\beta}}
du^\alpha \wedge dv^{\tilde{\beta}} ,
\label{fundS}
\end{equation}
and the line element takes the form~(\ref{lineel}), so that $\partial /\partial
u^\alpha$ and $\partial / \partial
\tilde{v}^{\tilde{\beta}}$ are null vectors. Any $N$-form
can be decomposed into a set of $(r,s)$ forms with $r$ factors of $du$ and
$s$ factors of $dv$ with $r+s=N$. The exterior derivative decomposes as
$d=\partial_u +\partial_v $ where $\partial_u : H^{(r,s)} \rightarrow
H^{(r+1,s)}$ and $\partial_v : H^{(r,s)} \rightarrow H^{(r,s+1)}$.

Consider first the case in which there is no torsion, $H=0$. Then the
conditions~(\ref{Scovconst}) and~(\ref{Sherm}) imply that the
geometry
is given in terms of some locally defined
scalar potential $\tilde{K}$, and the metric takes the form~(\ref{Kahl}) in
adapted coordinates; the sigma model with this geometry will be considered
further in the next section.

If $H\neq 0$, then the conditions~(\ref{Scovconst}) and~(\ref{Sherm}) imply
that the torsion
three-form is given in terms of the fundamental
   two-form~(\ref{fundS})
by
\begin{equation}
H = \left( \partial_u -\partial_v \right) S .
\end{equation}
The condition $dH=0$ then implies
\begin{equation}
\partial_u \partial_v S = 0
\end{equation}
so that locally there is a (1,0) form $k=k_\alpha du^\alpha$ and a (0,1)
form $\tilde{k} = \tilde{k}_{\tilde{\beta}}dv^{\tilde{\beta}}$ such that
\begin{equation}
S = \partial_u \tilde{k} + \partial_v k .
\end{equation}
The potentials $k,\tilde{k}$ are independent real 1-forms. The metric and
torsion
potential are   given, in
a suitable gauge, by eq.~(\ref{realgeom2}),
so that
\begin{equation}
H = \partial_u \partial_v \left( k + \tilde{k} \right) .
\label{realgeom3}
\end{equation}
If the condition~(\ref{condnoH}) holds for some locally defined potentials
$\kappa$, $\tilde{\kappa}$,   then the torsion vanishes and
\begin{equation}
S = \partial_u \partial_v \left( \tilde{\kappa} -\kappa \right) ,
\end{equation}
so that~(\ref{Kahl}) is satisfied with potential $\tilde{K}=\tilde{\kappa}
-\kappa$. We thus recover the results obtained from extended superspace; the
extended superspace appproach   gives
the general solution to the geometric constraints
immediately, without having to integrate differential equations.

If $H=0$, then the curvature two-form is a (1,1) form and the only
non-vanishing
components of the curvature are $R_{\alpha \tb \gamma \td}$. It follows that
the
Ricci tensor $R_{\alpha \tb}$ is proportional to $\tilde{C}_{\alpha \tb}$
and is given by
\begin{equation}
R_{\alpha \tb} = \partial_\alpha \partial_{\tb} \log | \det g_{\gamma
\td}|
\end{equation}
with $R_{\alpha \beta} = 0$. Thus the Einstein equation $R_{ij}=0$
is equivalent to demanding $SL(d,\R)$ holonomy and gives, with a
suitable choice of coordinates,
\begin{equation}
| \det g_{\gamma \td}| =1
\end{equation}
which is a Monge-Amp\`{e}re equation for $\tK$,
\begin{equation}
\det \left| \frac{\partial^2}{\partial u^\alpha \partial v^{\tb}} \tK
\right|
=1.
\end{equation}

If $H\neq 0$, then the metric and torsion are
preserved by the gauge transformations
\begin{equation}
\delta k_\alpha =\partial_\alpha \chi +\theta_\alpha , \ \  \ \ \delta
\tilde{k}_{\tilde{\alpha}} =-\partial_{\tilde{\alpha}}\chi +
\tilde{\theta}_{\tilde{\alpha}} ,
\label{gauge}
\end{equation}
where $\partial_{\tb} \theta_\alpha = \partial_\beta
\tilde{\theta}_{\tilde{\alpha}} =0$. In analogy with
eqs.~(\ref{w}), (\ref{U1R}), (\ref{U1gamm}), it is useful
to define the vector
\begin{equation}
\tilde{w}_i = H_{ijk} S^{jk}
\label{wtild}
\end{equation}
together with the $GL(1,\R)$ part of the curvature
\begin{equation}
\tilde{C}^{(+)}_{ij} = S^l{}_k R^{(+)k}_{lij}
\label{U1Rtild}
\end{equation}
and the $GL(1)$ part of the connection~(\ref{gamm}),
\begin{equation}
\tilde{\Gamma}^{(+)}_{i} = S^k{}_j \Gamma^{(+)j}_{ik} =
\Gamma^{(+)\alpha}_{i\alpha} -\Gamma^{(+)\tilde{\alpha}}_{i\tilde{\alpha}} .
\label{Gl1}
\end{equation}
For Kleinian signature $(n,n)$, the  holonomy of the connection $\Gamma^{(+)}$
is contained in $GL(n,\R)$. It will be contained in $SL(n,\R)$ if in
addition $\tilde{C}_{ij}^{(+)} =0$.

If $H\neq 0$, the condition~(\ref{Gam0}) of the complex case is replaced by
\begin{equation}
\tilde{\Gamma}^{(+)}_i =0
\label{tGam0}
\end{equation}
and this  implies that the one-loop field equation~(\ref{oneloop}) is
satisfied, provided the dilaton is chosen as in~(\ref{dilch}). Furthermore,
the condition~(\ref{tGam0}) implies $\tilde{C}^{(+)}_{ij}=0$ and so the
holonomy is in $SL(n,\R)$.

The field equation~(\ref{tGam0}) can
be obtained from the action~(\ref{HKMr}), but where now the metric is
given by~(\ref{realgeom1}) in terms of the potentials $k$, $\tk$ corresponding
to the real structure $S$, and it is these that are varied to give the field
equation~(\ref{tGam0}).

The real 1-form potentials $k$, $\tk$ can be dualised in the same way as
in the complex case to obtain a new form of the dual action as well as the
dual of the real geometry presented above. The first step is to  add
to~(\ref{HKM})
a Lagrange multiplier term of the form
\begin{equation}
\12 \tL^{\alpha \tb} \left( g_{\alpha \tb} -\partial_\alpha \tk_{\tb}
-\partial_{\tb} k_\alpha \right) .
\end{equation}
Eliminating $\tL^{\alpha \tb}$ from the resulting action, one recovers the
action~(\ref{HKM}) subject to the constraint $g_{\alpha \tilde{\beta}}
=\partial_\alpha \tilde{k}_{\tilde{\beta}} +\partial_{\tilde{\beta}} k_\alpha$.
Integrating
over the
vectors $k_\alpha$, $\tk_{\tb}$ instead yields the constraints
\begin{eqnarray}
\partial_\alpha  \tL^{\alpha \tb} & = & 0 \nonumber \\ \partial_{\tb}
\tL^{\alpha \tb} & = & 0
\end{eqnarray}
which in four dimensions are solved locally in terms of a scalar $\tilde{K}$
by
\begin{equation}
\tL^{\alpha \tb} = \tilde{L}^{\alpha \tb} ,
\label{solconstr}
\end{equation}
where $\tilde{L}^{\alpha \tb}$ is the \lq field strength' of $\tilde{K}$  given
by
\begin{equation}
\tilde{L}^{\alpha \tb} \equiv \epsilon^{\alpha \gamma \tb \tilde{\delta}}
\partial_\gamma \tilde{\partial}_{\tilde{\delta}} \tK  \label{defL}
\end{equation}
and $\epsilon^{\alpha \gamma \tb \tilde{\delta}}$ is the
antisymmetric tensor density (with $\epsilon^{1 \tilde{1} 2  \tilde{2 }}=1$).

The solution~(\ref{defL}) implies that the pseudo-K\"{a}hler metric $G_{\alpha \tb}=\partial_\alpha \partial_{\tb} \tilde{K}$ satisfies the
constraint
\begin{equation}
\det G_{\alpha \tb} = -1 
\end{equation}
for signature (2,2), or $\det G_{\alpha \tb} = +1$ for signature (4,0). Writing $G_{\alpha \tb} = \eta_{\alpha \tb} + \partial_\alpha \partial_{\tb} \tphi$ where $\eta_{\alpha \tb}$ is a flat background metric, the analysis
of ref.~\cite{AH3} then leads to the dual $D=4$ action
\begin{equation}
\int \partial \tphi \tdelbar \tphi +\frac{1}{3!} \tphi \tdelbar \tphi
\wedge \partial \tdelbar \tphi + \int \sqrt{G} G^{\alpha \tb} \partial_\alpha
\tilde{\Omega} \partial_{\tb} \tilde{\Omega} 
\end{equation}
for some scalar $\tilde{\Omega}$. Thus the dual geometry in four dimensions is a real form of self-dual gravity
without torsion determined by the potential $\tilde{K}$ coupled to the harmonic
scalar $\tilde{\Omega}$. The generalisation
to dimensions $D>4$ is straightforward, and the results are analogous to those
obtained in~\cite{AH3} for the complex case.

\section{(2,2) Supersymmetric Sigma Models}

\label{sect:22}

If the (2,2) supersymmetry closes off-shell, the sigma-model can be formulated
in terms of
off-shell (2,2) superfields.
For the usual untwisted (2,2) supersymmetry,
we introduce the complex superspace coordinates $z^\alpha$ ($\alpha =1,\ldots ,
{d_1} $),
$\theta_+$, $\theta_-$ together with the supersymmetry generators and
supercovariant derivatives
\begin{equation}
Q_+  =  \frac{\partial}{\partial \theta^+} -i\thetbar^+ \partial_{+} , \ \ Q_-
= \frac{\partial}{\partial \theta^-} -i\thetbar^-
\partial_{-} ,
\ee
and
\be
D_+  =  \frac{\partial}{\partial \theta^+} +i\thetbar^+ \partial_{+} , \ \ D_-
= \frac{\partial}{\partial \theta^-}
+i\thetbar^- \partial_{-} .
\ee
One can either introduce chiral superfields $U^\alpha$, $\Ubar^{\bbar}$
($\alpha , \bbar =1,\ldots
, d_1$) satisfying
\be
\Dbar_\pm U^{\alpha} =0, \ \ \ \ D_\pm \Ubar^{\bbar} =0,
\ee
or twisted chiral superfields $V^i$, $\Vbar^{\overline{j}}$ ($i, \jj =1,\ldots
d_2$) satisfying the constraints
\be
D_+ V^i =0, \ \ \ \ \Dbar_- V^i =0, \ \ \ \ D_- \Vbar^{\jj}=0, \ \ \  \ \Dbar_+
\Vbar^{\jj} =0.
\ee
The action for the K\"{a}hler sigma model is
\begin{equation}
S= \int d^2 \sigma d^4 \theta K(U, \Ubar ) ,
\label{actionK}
\end{equation}
where $K$ is the K\"{a}hler potential, so that the metric is given by
eq.~(\ref{kahlermetric}). The action and metric are invariant under the
K\"{a}hler gauge transformations
\begin{equation}
\delta K = f(U) +\overline{f}(\Ubar ) .
\end{equation}

The action~\cite{GHR}
\be
S= \int d^2 \sigma d^4 \theta K(U, \Ubar ,V, \Vbar )
\label{chtch}
\ee
defines a supersymmetric non-linear sigma model with torsion on a target space
of complex dimension $d_1 +d_2$ with coordinates $x^\mu = (u, \ubar ,v,\vbar
)$,
where $u$, $\ubar$, $v$ and $\vbar$ are the lowest components of the supefields
$U$, $\Ubar$, $V$ and $\Vbar$. The action~(\ref{chtch}) is invariant under
generalised K\"{a}hler gauge transformations
\begin{equation}
\delta K = f_1 (U,V) + f_2 (U,\Vbar ) +\overline{f}_1 (\Ubar ,\Vbar ) +
\overline{f}_2 (\Ubar , V) .
\end{equation}

The bosonic part of the component sigma model action is
\begin{equation}
S= \12 \int d^2 \sigma \left( g_{\mu \nu} \partial_\alpha x^\mu \partial^\alpha
x^\nu +b_{\mu \nu} \epsilon^{\alpha \beta} \partial_\alpha x^\mu \partial_\beta
x^\nu \right)
\label{fromN2}
\ee
where the metric $g_{\mu \nu}$ and the torsion potential $b_{\mu \nu}$ are
given by
\begin{eqnarray}
g_{\alpha \bbar} & = & K_{\alpha \bbar}, \ \ \ \ g_{i\jj} = -K_{i\jj} \nonumber
\\ b_{\alpha
\jj} & = & K_{\alpha \jj} , \ \ \ \ b_{i\bbar} = K_{i\bbar} .
\label{met1}
\end{eqnarray}
All other components of $g_{\mu \nu}$ and $b_{\mu \nu}$ not related to these by
complex conjugation or symmetry vanish, and $K_{\mu \nu \ldots \rho}$ denotes
the partial derivative $\partial_\mu \partial_\nu \ldots \partial_\rho K$. The
geometry is that of a Hermitean locally product space with two commuting
complex structures $J^{(\pm)\mu}_\nu$.
In the special case in which either $d_1 =0$ or $d_2 =0$, the torsion vanishes
and the target space
is K\"{a}hler.

For twisted (2,2) supersymmetry with the superalgebra~(\ref{susy}),
we introduce the real superspace coordinates $z^\mu$, $\theta_+$, $\ttheta_+$,
$\theta_-$, $\ttheta_-$ together with the supersymmetry generators and
supercovariant derivatives
\begin{eqnarray}
Q_+  & = &   \frac{\partial}{\partial \theta^+} -\ttheta^{+} \partial_{+} , \ \
Q_- = \frac{\partial}{\partial \theta^-}
-\ttheta^{-}
\partial_{-}, \nonumber \\ \tilde{Q}_+  & = &   \frac{\partial}{\partial
\ttheta^{+}} -\theta^+ \partial_{+} , \ \
\tilde{Q}_- = \frac{\partial}{\partial \ttheta^{-}} -\theta^-
\partial_{-},
\end{eqnarray}
and
\begin{eqnarray}
D_+  & = & \frac{\partial}{\partial \theta^+} +\ttheta^{+} \partial_{+} , \ \
D_- = \frac{\partial}{\partial \theta^-}
+\ttheta^{-}
\partial_{-}, \nonumber \\ \tilde{D}_+  & = & \frac{\partial}{\partial
\ttheta^{+}} +\theta^+ \partial_{+} , \ \
\tilde{D}_- = \frac{\partial}{\partial \ttheta^{-}} +\theta^-
\partial_{-} .
\end{eqnarray}
One can either introduce  superfields $U^{\alpha}$, $\tilde{U}^{\tilde{\beta}}$
satisfying the constraints
\begin{eqnarray}
D_+ \tilde{U}^{\tilde{\beta}} & = & 0, \ \ \ \ \tilde{D}_-  U^\alpha = 0,
\nonumber \\ \tilde{D}_+
U^\alpha &
= & 0, \ \ \ \ D_- \tilde{U}^{\tilde{\beta}} = 0,
\end{eqnarray}
or superfields $V^i$, $\tilde{V}^{\tilde{j}}$ satisfying the twisted constraints
\begin{eqnarray}
D_+ \tilde{V}^{\tilde{j}} & = & 0, \ \ \ \ \tilde{D}_-  \tilde{V}^{\tilde{j}} =
0, \nonumber \\ \tilde{D}_+
V^i &
= & 0, \ \ \ \ D_- V^i = 0.
\end{eqnarray}
The  pseudo-K\"{a}hler sigma model action is then
\begin{equation}
S= \int d^2 x d^2 \theta d^2 \ttheta \tilde{K} (U,\tilde{U}).
\end{equation}
The action, metric and torsion are  left invariant under the pseudo-K\"{a}hler
transformations
\begin{equation}
\delta \tilde{K} =f(U) +\tilde{f} (\tilde{U}) .
\end{equation}

The action
\be
S= \int d^2 \sigma d^2 \theta d^2 \ttheta \tilde{K}(U, \tilde{U} ,V, \tilde{V}
)
\label{rchtch}
\ee
defines a supersymmetric non-linear sigma model with torsion on a target space
of  dimension $2 (d_1 +d_2 )$ with coordinates $x^\mu = (u,
\tilde{u} ,v,\tilde{v} )$, where $u$, $\tilde{u}$, $v$ and $\tilde{v}$ are the
lowest components of the supefields
$U$, $\tilde{U}$, $V$ and $\tilde{V}$. The action~(\ref{rchtch}) is invariant
under
generalised pseudo-K\"{a}hler gauge transformations
\begin{equation}
\delta K = f_1 (U,V) + f_2 (U,\tilde{V} ) +\tilde{f}_1 (\tilde{U} ,\tilde{V} )
+
\tilde{f}_2 (\tilde{U} , V) .
\end{equation}

The bosonic part of the component sigma model action is
again given by~(\ref{fromN2}),
where the metric $g_{\mu \nu}$ and the torsion potential $b_{\mu \nu}$ are
given by
\begin{eqnarray}
g_{\alpha \tb} & = & K_{\alpha \tb}, \ \ \ \ g_{i\tilde{j}} = -K_{i\tilde{j}}
\nonumber \\ b_{\alpha
\tilde{j}} & = & K_{\alpha \tilde{j}} , \ \ \ \ b_{i\tb} = K_{i\tb} .
\label{met2}
\end{eqnarray}
All other components of $g_{\mu \nu}$ and $b_{\mu \nu}$ not related
to these by \lq real' conjugation or symmetry vanish. The geometry is
that of a real locally product space with two commuting real structures
$S^{(\pm)\mu}_\nu$; see~\cite{GHR} for details. In the special
case in which either $d_1 =0$ or $d_2 =0$, the torsion vanishes and
the target space is pseudo-K\"{a}hler.

The metrics and torsion potentials~(\ref{kahlermetric}), (\ref{met1})
and~(\ref{met2}) will
define consistent string backgrounds if the corresponding sigma-model is
conformally invariant. For the K\"{a}hler model, this will be the case if the
metric is Ricci-flat or equivalently if the curvature is self-dual (or
anti-self-dual), i.~e.\
\begin{equation}
\star R_{\mu \nu \rho \sigma} = \frac{1}{2} \epsilon_{\mu \nu}{}^{\lambda
\tau}{}
R_{\lambda \tau \rho \sigma} = \pm R_{\mu \nu \rho \sigma} .
\end{equation}
There are also generalisations of these self-dual solutions to the condition
for
one-loop conformal invariance with non-trivial dilaton, some of which were
discussed in ref.~\cite{KKL}.

The sigma model with action~(\ref{chtch}) was shown in~\cite{H2} to be one-loop
conformally invariant provided the $U(1)$ parts of the two curvature tensors
$R^{(\pm )}_{\mu \nu \rho \sigma}$ vanish,
\begin{equation}
C^{(\pm )}_{\mu \nu} =0 ,
\end{equation}
so that both connections $\Gamma^{(\pm )}$ have $SU (d_1 + d_2 )$ holonomy and
the
first Chern class vanishes; see~\cite{Htor} for a discussion of higher loops.
For the twisted case with action~(\ref{chtch}), the condition for one-loop
conformal
invariance is that
both connections $\Gamma^{(\pm )}$ have $SU (d_1, d_2 )$ holonomy.

Similarly, for the pseudo-K\"{a}hler sigma model with action~(\ref{rchtch}),
one-loop
conformal invariance will hold provided the $GL(1,\R)$ parts of the the two
curvature tensors (defined as in eq.~(\ref{U1Rtild})) vanish,
\begin{equation}
\tilde{C}^{(\pm )}_{\mu \nu} =0 .
\end{equation}
If this condition holds, then both connections $\Gamma^{(\pm )}$ will have
$SL(d_1
+d_2 ,\R)$ holonomy.

In refs.~\cite{ST,GMST}, it was argued that all sigma-models with the usual
(2,2)
supersymmetry can be formulated in superspace using chiral, twisted chiral and
semi-chiral~\cite{BLR} superfields. Semi-chiral superfields have twice as many
components as chiral or twisted chiral ones, half of which are auxiliary. Here
we
note that a real analogue of the semi-chirality condition can be imposed, viz.\
\be
D_+ W^\alpha =0, \ \ \ \ \tilde{D}_+ \tilde{W}^{\tilde{\beta}} =0, \ \ \ \ D_-
\tilde{X}^{\tilde{j}}=0, \ \ \  \ \tilde{D}_-
\tilde{X}^{\tilde{j}} =0.
\ee
This leads to a straightforward generalisation of many of the
results  of~\cite{ST,GMST}   to   twisted (2,2) supersymmetric theories.

\newpage

\section{Summary and Discussion}

\label{sect:twistcomm}

To summarise, the usual supersymmetry algebra of type $(p,q)$ can be
generalised to include the
possibility of twisted heterotic supersymmetry, as in~(\ref{susy})
and~(\ref{eta}), and a superspace for this can be defined. The geometry of the
heterotic sigma models which realise this algebra is a generalisation of
K\"{a}hler geometry with torsion,
or a further generalisation involving real structures squaring to +1.

A superspace formulation of the
supersymmetric nonlinear
sigma models with untwisted or twisted $(p,q)$ supersymmetry was given in
section~\ref{sect3} in a formalism in which (1,1) supersymmetry is manifest.
For such sigma models,
more general isometries of the form~(\ref{rigidtransf}) can be considered,
where the vectors $\xi_a$ are
Killing vectors which are  holomorphic
with respect to each complex structure, or \lq holomorphic' in a generalised
sense with respect to each real
structure. The gauging of such isometries can be obtained from a
straightforward extension of the results
of refs.~\cite{HP1,HP2,HSP}.

The
results concerning the amount and
type of supersymmetry that can be realised can be summarised in terms of the
holonomy group of the
connection with torsion. The various possibilities, which depend on the
signature of the target space, are
listed in table 1.

\begin{center}
\vspace{.5cm}

\begin{tabular}{||c|c|c|c||}  \hline
{\em Target Signature} & {\em Holonomy of $\Gamma^{(+)}$} & {\em
Geometry when Torsion-Free} & {\em Supersymmetry}
\\ \hline $(d_1,d_2)$ & $O(d_1,d_2)$ & no restriction &
(1,1) \\ $(2n_1 ,2n_2 )$ & $U(n_1 , n_2 )$ &
K\"{a}hler  & (2,1)
\\ $(4m_1,4m_2)$ &
$USp(2m_1,2m_2)$ & hyper-K\"{a}hler & (4,1)
\\    $(2n,2n)$ &
$GL(n,\R)$ & pseudo-K\"{a}hler & twisted (2,1) \\ $(4m,4m)$ & $Sp(2m,\R)$ &
pseudo-hyper-K\"{a}hler &
twisted (4,1) \\ \hline
\end{tabular}

\vspace{.2cm}

\end{center}

{\bf Table 1:} \begin{quote}
The relation of right-handed supersymmetry to the holonomy of the connection
with
torsion $\Gamma^{(+)}$. We give the type of
  target space geometry for the case in which the torsion vanishes.
\end{quote}

\vspace{.3cm}

For example, in the case of target spaces of Kleinian signature $(d,d)$
with a single real structure, the holonomy group is contained in $GL(d,\R)$ and
the model has twisted (2,1)
supersymmetry. The geometry generalises that of the usual (2,1) sigma model: in
particular, the metric and
torsion potential are given by~(\ref{realgeom1}), (\ref{realgeom2}) where $k$
and $\tilde{k}$ are {\em
independent} real forms. This model can be formulated in superspace as shown in
section~\ref{sect:t21}. Sigma models with untwisted or twisted $N=2$
supersymmetry can also be formulated in superspace, and this leads to new
pseudo-K\"{a}hler (without torsion) and twisted pseudo-K\"{a}hler (with
torsion) sigma models whose geometry is determined by a scalar potential
analogous to the twisted K\"{a}hler potential of
ref.~\cite{GHR}. If the torsion vanishes, then the twisted (2,1) supersymmetric
model
reduces to the pseudo-K\"{a}hler model. These real models are listed
in table 2.
\begin{center}
\vspace{.5cm}

\begin{tabular}{||c|c|c||}  \hline
{\em Target Geometry} & {\em Superfields} & {\em Supersymmetry} \\ \hline  real
with
torsion & $U_{(2,1)}, \tilde{V}_{(2,1)}$ & twisted (2,1)  \\ pseudo-K\"{a}hler
& $U_{(2,2)},
\tilde{U}_{(2,2)}$ & twisted (2,2)  \\ twisted-pseudo-K\"{a}hler & $U_{(2,2)},
\tilde{V}_{(2,2)}$ & twisted (2,2) \\ \hline
\end{tabular}

\vspace{.2cm}

\end{center}

{\bf Table 2:} \begin{quote}
Geometry, superfields and supersymmetry of some sigma models with real
target spaces.
\end{quote}

\vspace{.3cm}

It is remarkable how much of the geometry based on a complex structure $J$
carries over to the case of a
real structure $S$. Instead of using complex numbers, it is useful to introduce
{\em double numbers} in
this context~\cite{BGPPR}. These are based on a real unit $e$ which satisfies
\begin{equation}
e^2 = +1
\label{e}
\end{equation}
instead of the usual imaginary unit $i$ satisfying $i^2 =-1$.
It is useful to define a real conjugation  taking $ e \to -e$, so that
$(x+ey)^*=x-ey$
for real numbers $x,y$.

For example, consider the formulation of the twisted (2,1) sigma model of
section~\ref{sect:t21} using
double numbers. The real structure $S^i{}_j$  takes the form
\begin{equation}
S^i{}_j = e\left( \begin{array}{cc} \delta^{\alpha}_{\beta} & 0 \\ 0 &
-\delta^{\tilde{\alpha}}_{\tilde{\beta}} \end{array} \right)
\end{equation}
in an adapted coordinate system. The fundamental two-form is then
\begin{equation}
S = \frac{1}{2} S_{ij}d\phi^i \wedge d\phi^j = -e g_{\alpha \tilde{\beta}}
du^\alpha \wedge dv^{\tilde{\beta}} .
\end{equation}
If $H\neq 0$, the torsion is given in terms of the fundamental two form by
\begin{equation}
H = e \left( \partial_u -\partial_v \right) S .
\end{equation}
The closure of $H$ then implies
\begin{equation}
e\partial_u \partial_v S = 0
\end{equation}
and the geometry is given, in a suitable gauge, by
eqs.~(\ref{realgeom1})-(\ref{realgeom3}). The metric
and torsion are preserved by the gauge transformations
\begin{equation}
\delta k_\alpha =e\partial_\alpha \chi +\theta_\alpha , \ \ \ \ \delta
\tilde{k}_{\tilde{\alpha}} =-e\partial_{\tilde{\alpha}}\chi +
\tilde{\theta}_{\tilde{\alpha}} ,
\end{equation}
where $\partial_{\tb} \theta_\alpha = \partial_\beta
\tilde{\theta}_{\tilde{\alpha}} =0$.
The superspace action is
\begin{equation}
S = -e\int d^2 \sigma d\theta^{+}d\tilde{\theta}^{+}d\theta^{-} \left(
k_\alpha D_- U^\alpha -\tilde{k}_{\tilde{\alpha}} D_-
\tilde{V}^{\tilde{\alpha}}
\right)
\label{lastS} ,
\end{equation}
where the superfields $U^\alpha$, $\tilde{V}^{\tilde{\alpha}}$ are chiral with
respect to the superderivatives
\begin{equation}
D_+ = \frac{\partial}{\partial \theta^{+}} + e\tilde{\theta}^{+}
\frac{\partial}{\partial \sigma^{+}}  , \ \ \ \ \tilde{D}_+ =
\frac{\partial}{\partial \tilde{\theta}^{+}} + e\theta^{+}
\frac{\partial}{\partial \sigma^{+}}.
\end{equation}

If the twisted (2,1) superspace action~(\ref{lastS}) is required to be real
self-conjugate with respect
to the   conjugation $e\to  {e}^*=-e$, i.~e.\ if
\begin{equation}
S =  {S}^*,
\label{self}
\end{equation}
then we find that the potentials $k$ and $\tilde{k}$ are real conjugates,
\begin{equation}
   { \tilde{k} }=k^* .
\label{depend}
\end{equation}
This is the generalisation to the double numbers of the reality condition
\begin{equation}
S= {S}^*
\label{realonS}
\end{equation}
on the
action~(\ref{I21real}), which implies that $k=(\overline{k})^*$; in turn, this
implies hermicity of the
metric and antihermicity of the
torsion potential given
in~(\ref{complgeom}). For the general models we have discussed, the
condition~(\ref{realonS}) does
not hold, the potentials $k$ and $\tilde{k}$ are independent and the action is
not
real self-conjugate.

Setting $e=1$, the formulations of previous sections are recovered, but
introducing $e$ is a useful
book-keeping device. In particular, it leads to the introduction of the real
conjugation operation,
and makes the structure similar to that of the complex case.

\vspace{1.5cm}

{\bf Acknowledgements}
\\
\\
        M.\ A.\  would like to thank the Theory Division at CERN, where parts of
this work were carried out, for hospitality and support.


\begin{thebibliography}{99}

\bibitem{H1} C.\ M.\ Hull, {\em Actions for (2,1) Sigma Models and Strings},
Nucl.\ Phys.\ {\bf B509} (1998) 252, [hep-th/9702067].

\bibitem{HW} C.\ M.\ Hull and E.\ Witten, {\em Supersymmetric Sigma Models and
the Heterotic String}, Phys.\ Lett.\ {\bf B160}
(1985) 398.

\bibitem{H3} C.\ M.\ Hull, {\em Complex Structures and Isometries in the (2,0)
Supersymmetric
Nonlinear Sigma Model}, Mod.\ Phys.\ Lett.\ {\bf 5} (1990) 1793.

\bibitem{H2} C.\ M.\ Hull, {\em Sigma Model Beta-Functions and String
Compactifications},
Nucl.\ Phys.\ {\bf B267} (1986) 266.

\bibitem{CMHnew} C.\ M.\ Hull, {\em Timelike T-Duality, de Sitter Space, Large
N Gauge Theories and Topological Field Theory}, JHEP {\bf 9807} (1998) 021,
[hep-th/9806146].

\bibitem{CMHnew2} C.\ M.\ Hull, {\em Duality and the Signature of Space-Time},
JHEP {\bf 9811} (1998) 017, [hep-th/9807127].

\bibitem{forza} M.\ Ademollo, L.\ Brink, A.\ D'Adda, R.\ D'Auria, E.\
    Napolitano, S.\ Sciuto,
E.\ Del Giudice, P.\ Di Vecchia, S.\ Ferrara,              F.\ Gliozzi, R.\
Musto, R.\ Pettorino and J.\ H.\
Schwarz, {\em Dual String with $U(1)$ Color Symmetry},
Nucl.\ Phys.\
{\bf B111} (1976) 77.

\bibitem{OV1} H.\ Ooguri and C.\ Vafa, {\em $N=2$ String Magic}, Mod.\ Phys.\
           Lett.\ {\bf A5}
(1990) 1389; \ {\em Geometry of $N=2$ Strings}, Nucl.\ Phys.\              {\bf
B361} (1991) 469.

\bibitem{OV2} H.\ Ooguri and C.\ Vafa,  {\em $N=2$ Heterotic Strings},
Nucl.\ Phys.\ {\bf B367} (1991)  83.

\bibitem{KM1} D.\ Kutasov and E.\ Martinec, {\em New Principles for
String/Membrane
Unification}, Nucl.\ Phys.\ {\bf B477} (1996) 652,
[hep-th/9602049].

\bibitem{KM2} D.\ Kutasov, E.\ Martinec and M.\ O'Loughlin, {\em Vacua of
M-Theory and $N=2$ Strings}, Nucl.\ Phys.\
{\bf B477} (1996) 675, [hep-th/9603116].

\bibitem{M1} E.\ Martinec, {\em Geometrical Structures of M-Theory},
[hep-th/9608017].

\bibitem{KM3} D.\ Kutasov and E.\ Martinec, {\em M-Branes and $N=2$ Strings},
Class.\ Quantum Grav.\ {\bf 14}
(1997) 2483, [hep-th/9612102].

\bibitem{M2} E.\ Martinec, Lectures at the 1997 Carg\`{e}se Summer School
{\em Strings, Branes and Dualities}, May 26 - June 14, 1997.

\bibitem{M3} E.\ Martinec, {\em Matrix Theory and $N=(2,1)$ Strings},
[hep-th/9706194].

\bibitem{M4} E.\ Martinec, {\em M-Theory and $N=2$ Strings}, [hep-th/9710122].

\bibitem{BFSS} T.\ Banks, W.\ Fischler, S.\ Shenker and L.\ Susskind, {\em M
Theory as a Matrix Model: A Conjecture}, Phys.\
Rev.\ {\bf D55} (1997) 5112, [hep-th/9610043].

\bibitem{AH1} M.\ Abou Zeid and C.\ M.\ Hull, {\em Geometry, Isometries
and Gauging of (2,1) Heterotic Sigma Models}, Phys.\ Lett.\ {\bf B398} (1997)
291, [hep-th/9612208].

\bibitem{AH2} M.\ Abou Zeid and C.\ M.\ Hull, {\em The Gauged (2,1)
Heterotic Sigma Model}, Nucl.\ Phys.\ {\bf B513} (1998) 490, [hep-th/9708047].

\bibitem{AH3} M.\ Abou Zeid and C.\ M.\ Hull, {\em Conformal Invariance and
Duality in Self-Dual Gravity and (2,1) Heterotic String Theory},
Phys.\ Lett.\ {\bf B419} (1998) 139, [hep-th/9710096]; {\em Erratum - ibid.},
{\bf B431} (1998) 459.

\bibitem{MAth} M.\ Abou Zeid, Ph.\ D.\ Thesis, London (1998).

\bibitem{KKL} E.\ Kiritsis, C.\ Kounnas and D.\ L\"{u}st, {\em A Large Class of
New Gravitational and Axionic Backgrounds for Four-dimensional Superstrings},
Int.\ J.\ Mod.\ Phys.\ {\bf A9} (1994) 1361, [hep-th/9308124].

\bibitem{GHR} S.\ J.\ Gates, Jr., C.\ M.\ Hull and M.\ Ro\v{c}ek, {\em Twisted
Multiplets and New
Supersymmetric Nonlinear Sigma Models}, Nucl.\
Phys.\ {\bf B248} (1984) 157.


\bibitem{H02} C.\ M.\ Hull, {\em Compactifications of the Heterotic
Superstring}, Phys.\ Lett.\ {\bf
B178} (1986) 357.






\bibitem{FN} A.\ Fr\"{o}hlicher and A.\ Nijenhuis, {\em Theory of Vector-valued
Differential Forms I: Derivations in the Graded Ring of Differential Forms},
Nederl.\ Acad.\
Wetensch.\ Proc.\ Ser.\ {\bf A59} (1956) 339.

\bibitem{NJH} N.\ J.\ Hitchin, {\em Hypersymplectic Quotients}, in Acta
Academie
Scientiarum Taurinensis, Supplemento al Numero 124 degli Atti della Accademia
delle Scienze di Torino
Classe di Scienze Fisiche, Matematiche e Naturali (1990).

\bibitem{BGPPR} J.\ Barret, G.\ W.\ Gibbons, M.\ J.\ Perry, C.\ N.\ Pope and
P.\ Ruback, {\em Kleinian Geometry and the $N=2$ Superstring}, Int.\ J.\ Mod.\
Phys.\ {\bf A9} (1994) 1457, [hep-th/9302073].

\bibitem{Gilmore} R.\ Gilmore, {\it Lie Groups, Lie ALgebras and Some of Their
Applications}, John Wiley \& Sons  (1974).

\bibitem{HP1} P.\ S.\ Howe and G.\ Papadopoulos, {\em Ultraviolet Behavior of
Two-Dimensional
Supersymmetric Nonlinear Sigma Models}, Nucl.\ Phys.\ {\bf B289}
(1987) 264.


\bibitem{HP2} P.\ S.\ Howe and G.\ Papadopoulos,
{\em Further Remarks on the Geometry of Two-Dimensional Nonlinear Sigma
Models},
Class.\ Quantum Grav.\
{\bf 5} (1988) 1647.

\bibitem{DS} M.\ Dine and N.\ Seiberg, {\em (2,0) Superspace}, Phys.\ Lett.\
{\bf B180} (1986) 364.

\bibitem{Buscher} T.\ H.\ Buscher, {\em Quantum Corrections and Extended
Supersymmetry in New $\sigma$-Models}, Phys.\ Lett.\ {\bf B159} (1985) 127.

\bibitem{Htor} C.\ M.\ Hull, {\em The Geometry of $N=2$ Strings with Torsion},
Phys.\ Lett.\ {\bf B387} (1996) 497, [hep-th/9606190].

\bibitem{HSP} C.\ M.\ Hull, G.\ Papadopoulos and W.\ Spence, {\em  Gauge
Symmetries for
$(p,q)$
Supersymmetric Sigma Models}, Nucl.\ Phys.\ {\bf B363} (1991) 593.

\bibitem{ST} A.\ Sevrin and J.\ Troost, {\em Off-Shell Formulation of $N=2$
Non-Linear Sigma Models}, Nucl.\ Phys.\ {\bf B492} (1997) 623,
[hep-th/9610102].

\bibitem{GMST} M.\ T.\ Grisaru, M.\ Massar, A.\ Sevrin and J.\ Troost, {\em The
Quantum Geometry of $N=(2,2)$ Non-Linear Sigma Models}, Phys.\ Lett.\ {\bf
B412} (1997) 53, [hep-th/9706218].

\bibitem{BLR} T.\ H.\ Buscher, U.\ Lindstr\"{o}m and M.\ Ro\v{c}ek, {\em New
Supersymmetric Sigma Models with Wess-Zumino Terms}, Phys.\ Lett.\
{\bf B202} (1988) 94.



\end{thebibliography}
\end{document}